\documentclass{article}
\usepackage{graphicx} 

\usepackage{natbib}
\usepackage{fullpage}

\title{Clarification of `Algorithmic Collusion without Threats'}
\author{Jason Hartline}
\date{}

\begin{document}

\maketitle

\begin{abstract}
  This brief note clarifies that the scenario described in
  \citet{ACKRZ-25}---titled `Algorithmic Collusion without
  Threats'---is not one of collusion, but one where one player is
  behaving non-competitively and the other is behaving competitively.
\end{abstract}

It is important for academic discourse at the interface between
computer science and law to identify definitions for concepts that
make sense both in the legal domain and in the computational domain.
{\em Algorithmic collusion}---where there is a concern
that algorithms might facilitate the coordination of sellers on
supra-competitive prices---is one such concept.  This brief note is to
point out that the phenomenon studied by \citet{ACKRZ-25} is not
algorithmic collusion.

\citet{ACKRZ-25}---titled `Algorithmic Collusion without
Threats'---shows that very simple strategies sustain supra-competitive
prices in Nash equilibrium of a simple repeated pricing game.  In this
equilibrium, one player plays by a static price distribution and the
other player plays a strategy that learns to best respond (at least)
to any static price distribution.  In this Nash equilibrium of the
repeated game, both players obtain a constant fraction of the monopoly
revenue, while the ``competitive'' Nash equilibrium of the stage game
has (approximately) no revenue for either player.  The authors claim
that both players' strategies are reasonable in that they are
optimizing against their opponent's strategy.  The second player is
reasonable for learning to best respond to the induced distribution of
the demand.  Given that the second player is doing this, the first
player is reasonable in responding to the second player in a way that
improves revenue.

While the outcome above is non-competitive, it is not a collusion.
Collusion fundamentally requires both players to be complicit. Only
the first player in this outcome is behaving non-competitively.
In fact, the informal argument that the first player is behaving
reasonably is flawed.

There is an important distinction in the study of repeated games
between equilibria of the stage game that are repeated and equilibria
of the repeated game as a whole.  Stage game equilibria are generally
a subset of a much larger set of repeated game equilibria.  The
economics literature is unambiguous that {\em competitive prices} are
ones that result from an equilibrium in the stage game.  Equilibria of
a repeated game that are not equilibria of the stage game are, by this
definition, {\em non-competitive}.  The folk theorem suggests that there
are many such non-competitive equilibria.  Some are from strategies
with threats like tit-for-tat and some---as obtained in
\citet{ACKRZ-25}---are not.  The former looks like collusion; the
latter does not, but both are non-competitive.  For both cases,
strategic optimization at the level of the repeated game is exactly
the ``unreasonable'' behavior that needs to be curtailed to enable
competitive prices.

Collusion is a special case of non-competitive behavior where, in the
legal theory, agreements are made to distort prices away from
competitive prices.  Recently, the algorithmic collusion literature
has also considered tacit collusion where, rather than sellers making
an agreement, the algorithms learn dynamics that sustain
supra-competitive prices.  In all of the legal and academic discourse
on collusion, it is fundamentally about the choices of two or more
parties that result in non-competitive prices.  An outcome can only be
collusion if choices of both players are implicable in arriving at
that outcome.  Specifically, in a collusion there must always be a
reasonable strategy that a player can make that is non-collusive.

Algorithms are used to price because market conditions such as the
demand and level of competitive prices are unknown.  A minimum
requirement to consider a pricing algorithm as good is, given a static
demand, i.e., the probability of sale at every price is the same
across rounds, the pricing algorithm learns to optimize with respect
to this demand.  There may be more requirements of good algorithms
when considering dynamic settings of price competition, e.g., see \citet{har-26},  but these
additional requirements cannot be distinguished when considering only
static demand.  Thus, for static
demand an algorithm is indistinguishable from a good algorithm if it
learns to best respond.

As any reasonable definition collusion requires that there be a strategy
that is non-collusive, and as algorithms that learn to best respond
against static demand are indistinguishable from good algorithms, an
algorithmically reasonable definition of collusion cannot call an
outcome collusion when all but one player is best responding to a
static induced demand.

While \citet{ACKRZ-25} is not a paper about collusion, it does have an
important message for the literature on pricing algorithms.  A
regulator may struggle to distinguish between sellers who na\"ively
price without optimizing and sellers who are manipulating the market,
like the first player in their example.  The lesson is that to ensure
competitive prices, both such sellers must be regulated against.  Such
regulation may be difficult in practice.

\bibliographystyle{apalike}
\bibliography{refs}

\begin{thebibliography}{}

\bibitem[Arunachaleswaran et~al., 2025]{ACKRZ-25}
Arunachaleswaran, E.~R., Collina, N., Kannan, S., Roth, A., and Ziani, J.
  (2025).
\newblock Algorithmic collusion without threats.
\newblock In {\em 16th Innovations in Theoretical Computer Science Conference
  (ITCS 2025)}, pages 10--1. Schloss Dagstuhl--Leibniz-Zentrum f{\"u}r
  Informatik.

\bibitem[Hartline, 2026]{har-26}
Hartline, J. (2026).
\newblock The economics of no-regret learning algorithms.
\newblock In {\em Advances in Economics and Econometrics: Thirteenth World
  Congress}, volume~2. Cambridge.

\end{thebibliography}

\end{document}